\documentclass{aa}
\usepackage{graphicx,epsfig,fancyhdr,psfig,rotating,amsmath,longtable}
\usepackage{natbib} 
\usepackage{txfonts}

\bibpunct{(}{)}{;}{a}{}{,} 
  \newcommand{\Msolar}{\mbox{\,$\rm M_{\odot}$}}        
  \newcommand{\Teff}{\mbox{\,\em T$_{\rm eff}$}}        
  \newcommand{\logg}{\mbox{\,log $g$}}                  

  \newcommand{\vsini}{\mbox{\,$v\,\sin i$}}              
  \newcommand{\ie}{\mbox{i.e.}}                          
  \newcommand{\eg}{\mbox{e.g.}}                          
  \newcommand{\etal}{\mbox{et~al.}}                      
  \newcommand{\ang}{\,\mbox{\AA}}                        
  \newcommand{\kmsec}{\,\mbox{$\mbox{km}\,\mbox{s}^{-1}$}} 
  \newcommand{\kelvin}{\,\mbox{K}}                       
  \def\simge{\mathrel{\raise1.16pt\hbox{$>$}\kern-7.0pt
    \lower3.06pt\hbox{{$\scriptstyle \sim$}}}}           
  \def\simle{\mathrel{\raise1.16pt\hbox{$<$}\kern-7.0pt
    \lower3.06pt\hbox{{$\scriptstyle \sim$}}}}           

\begin{document}

\title{Discovery of kilogauss magnetic fields in three DA white dwarfs
\thanks{Based on observations collected at the European Southern Observatory, 
Paranal, Chile, under programme ID 70.D-0259}}
\author{R. Aznar Cuadrado\inst{1} \and S. Jordan\inst{2, 3} 
        \and R. Napiwotzki\inst{4}  \and H. M. Schmid\inst{5} 
        \and S. K. Solanki\inst{1} \and G. Mathys\inst{6}
       }
\offprints{R. Aznar Cuadrado, \email{aznar@linmpi.mpg.de}}
\institute{Max-Planck-Institut f\"{u}r Aeronomie, Max-Planck-Str. 2, 
37191 Katlenburg-Lindau, Germany \and
Institut f\"{u}r Astronomie und Astrophysik, Eberhard-Karls-Universit\"{a}t 
T\"{u}bingen, Sand 1, 72076 T\"{u}bingen, Germany \and
Astronomisches Rechen-Institut, M\"{o}nchhofstr. 12-14, 69120 Heidelberg, 
Germany \and
Department of Physics \& Astronomy, University of Leicester, University 
Road, Leicester LE1 7RH, UK \and
Institut f\"{u}r Astronomie, ETH Zentrum, 8092 Z\"{u}rich, Switzerland \and
European Southern Observatory, Casilla 19001, Santiago 19, Chile
}

\date{}

\abstract{
We have detected longitudinal magnetic fields between 2 and 4\,kG in three 
(WD\,0446$-$790, WD\,1105$-$048, WD\,2359$-$434) out of a sample of 12 normal 
DA white dwarfs by using optical spectropolarimetry done with the VLT Antu 
8 m telescope equipped with FORS1. With the exception of 40 Eri B (4\,kG) 
these are the first positive detections of magnetic fields in white dwarfs 
below 30\,kG. Although suspected, it was not clear whether a significant 
fraction of white dwarfs contain magnetic fields at this level. These fields 
may be explained as fossil relics from magnetic fields in the main-sequence 
progenitors considerably enhanced by magnetic flux conservation during the 
shrinkage of the core. A detection rate of 25 \% (3/12) may indicate now for 
the first time that a substantial fraction of white dwarfs have a weak 
magnetic field. This result, if confirmed by future observations, would form a 
cornerstone for our understanding on the evolution of stellar magnetic fields.
}

\authorrunning{Aznar Cuadrado \etal}
\titlerunning{Discovery of kilogauss magnetic fields in three WDs}

\maketitle

\section{Introduction}
The major goal of this work is a better understanding of the role played by 
magnetic fields in the formation and evolution of stars. Magnetic fields are 
already an important ingredient during the collapse and fragmentation of 
protostellar clouds which ultimately determines the initial field of pre-main 
sequence stars. On the main sequence and at later stages of evolution the 
magnetic fields have a major impact on the angular momentum loss and stellar 
winds, on building-up chemical anomalies and abundance inhomogeneities across 
the stellar surface, on convection and the related coronal activity, and other 
evolutionary processes especially in interacting binaries. Very strong 
magnetic fields are detected in several white dwarfs and always present in 
pulsars \citep{Mestel:01}. The first detection of a magnetic field on a white 
dwarf was made by \cite{Kemp-etal:70} on Grw+$70^\circ$ 8247, and large 
spectroscopic and polarimetric surveys have been carried out in the last two 
decades \citep[\eg\ ][]{Hagen-etal:87, Reimers-etal:94, Schmidt-Smith:95, 
Putney:95, Kawka-etal:03}. For the white dwarfs the magnetic fields could 
simply be ``fossil'' remnants of the fields already present in main-sequence 
stars, but strongly amplified by contraction. This hypothesis assumes that the 
magnetic flux (\eg\ through the magnetic equator) is conserved to a large 
extent during the stellar evolution.

According to the field amplification theory, the white dwarfs play an 
important role in the investigation of stellar magnetic fields. In 
main-sequence stars magnetic fields  have been detected directly mainly 
for 
peculiar magnetic Ap and Bp stars with rather well organized fields and field 
strengths of the order $10^2-10^4$ Gauss. For weak fields in A to O stars 
($B<10^2$ G) direct magnetic field detections are still very rare 
\citep[\eg\ the field detections in early B stars reported by][]
{Neiner-etal:03a,Neiner-etal:03b}.
For sun-like stars ample evidence (coronal activity) for the presence of 
complicated small-scale fields exists, but direct measurements are only 
possible for the more active stars 
\citep{Saar:96,Ruedi-etal:97,Valenti-Johns:01}.
The contraction to a white dwarf amplifies the magnetic fields by about 4 
orders of magnitude, so that weak and often undetectable magnetic fields on 
the main sequence become measurable during the white dwarf phase. This is 
supported by the known magnetic white dwarfs with megagauss fields 
($B=10^6-10^9$~G). Their frequency and space distribution, as well as 
their mass, are consistent with the widely accepted view that they are 
the descendents of the magnetic Ap and Bp stars \citep[\eg\ ][]{Mathys:01}.
Another origin  seems to be required for the magnetic degenerates with weaker 
fields (unless magnetic flux is lost during the contraction phase). Magnetic 
main-sequence stars with weaker magnetic fields have been suggested as their 
possible progenitor candidates \citep{Schmidt-etal:03,Kawka-etal:03}.
The B stars on which weaker fields have been detected may be the missing 
stars. However, even the most sensitive observations are limited to some tens 
of gauss on main-sequence stars.

Thus, magnetic field amplification during stellar evolution may offer the 
opportunity to investigate $\sim 1$~G magnetic fields (averaged global fields) 
in normal main-sequence stars with observations of $\sim 1$~kG magnetic fields 
during the white dwarf stage.   
White dwarfs with magnetic fields below 100\,kG have been either found by 
searching for circular polarization \citep{Schmidt-Smith:94} or by looking for 
Zeeman splitting of narrow NLTE line cores in the Balmer lines, particularly  
in H$\alpha$ \citep{Koester-etal:98}.
However, the splitting becomes undetectable in intensity spectra for weak 
fields ($<20$ kG) or for objects without narrow line core. Therefore, 
spectropolarimetry is the most promising technique for successful detections 
of weak magnetic fields. 
Up to now detections of magnetic fields below 30 kG have not been achieved, 
except for the very bright white dwarf 40 Eri B ($V=8.5$), in which 
\cite{Fabrika-etal:03} have detected a magnetic field as low as 4\,kG.
The magnetic field detection limit can now be pushed down to a few kG 
for many white dwarfs with spectropolarimetry using 8-10~m class
telescopes.

In this paper we present and analyse VLT spectropolarimetric data of a sample 
of 12 white dwarfs in a search for weak magnetic fields. In Sect. 2 the 
observations and data reduction are described, while in Sect. 3 the 
observational method for obtaining the Stokes parameter ($V/I$) is described. 
Sect. 4 presents the method for determining weak magnetic fields analysing the 
circular polarisation due to a given magnetic field. In that section we also 
present the results of our analysis, along with the description of the  
$\chi^2$-minimisation procedure applied to our data. The determination of the 
atmospheric and stellar parameters is presented in Sect. 5 and compared with 
those found in the literature. A discussion and conclusions are presented in 
Sect. 6. 

\section{Observations and data reduction}
Spectropolarimetric observations of a sample of 12 normal DA white dwarfs were 
carried out during the period 4 November 2002 -- 3 March 2003, in service mode 
by ESO staff members using FORS1 at the 8 m Unit Telescope 1 (UT1) of the Very 
Large Telescope, ``Antu''. FORS1 is a multi-mode focal reducer imager and 
grism spectrograph equipped with a Wollaston prism and rotatable retarder 
plate mosaics in the parallel beam allowing linear and circular polarimetry 
and spectropolarimetry \citep{Appenzeller-etal:98}.

With a 0.8$\arcsec$ wide slit we obtained a (FWHM) spectral resolution of 
4.5 \ang. The data were recorded with a backside-illuminated thinned, AR 
coated Tek. 2048$\times$2048 CCD with 24 $\mu$m pixels which correspond to a 
pixel scale of 0.2$\arcsec$/pixel in spatial and 1 \ang/pixel in spectral 
direction. 

Spectra were acquired with grism G600B, which allows observations in the 
spectral range 3400--6000 \ang, covering all H\,{\sc i} Balmer lines 
from H$\beta$ to the Balmer jump simultaneously. 
A reflex image from the FORS optics affects the wavelength
region from 4000 to 4100~{\AA}\ which corresponds to the blue wing 
of the H$\delta$ line. Although the reflex shows up in the intensity
spectrum it seems to produce no spurious signal in circular polarisation. 
The reflex is known to occur for this particular grism G600B / Wollaston 
configuration \citep[\eg\ ][]{Schmid-etal:03}. 

The observations were split into several cycles to avoid saturation. 
The feasibility of circular spectropolarimetry with the required high 
signal-to-noise ratio using FORS1 had been demonstrated by 
 \cite{Bagnulo-etal:02,Bagnulo-etal:04} with test measurements of one 
magnetic 
and one non-magnetic A-star. Stellar rotation may cause continuous changes in 
the field orientation and, therefore, in the polarisation signal. Hence, we 
have split the observations of an individual target into more than one 
polarisation measurement taken during different nights (see Sect. 3). 
In this way, there is a much lower probability that a candidate with a 
magnetic field escapes detection due to a special orientation at the time of 
the observations (where circular polarisation cancels).
Hence, having for most objects data for more than one epoch, we are able to 
assess for the presence of rotational modulation of a possible magnetic field.

\subsection{Selection of the sample}
We have searched for the circular polarisation signatures of the Zeeman effect 
in the Balmer lines in 12 of the  brightest southern white dwarfs (having 
$11.4 ~{\rm mag} < V < 14 ~{\rm mag}$)  with RA between 0$^{\rm h}$ and 
12$^{\rm h}$. 

Our sample was carefully selected on the basis of VLT-UVES spectra taken 
within the SPY survey \citep{Napi-etal:03}: SPY ({\bf S}upernovae type Ia 
{\bf P}rogenitor surve{\bf Y}) is a radial velocity search for close binary 
systems of two white dwarfs (double degenerates; DD). If these systems are 
close enough they will merge due to gravitational wave radiation and if the 
combined mass of these mergers exceed the Chandrasekhar limit for white dwarfs 
these are potential progenitors of Supernovae type Ia. 

SPY was carried out with the high-resolution Echelle spectrograph UVES at the 
Kueyen (UT2) of VLT. With the set-up used for SPY UVES provides almost 
complete spectral coverage of the wavelength range from 3200\,{\AA}\ to 
6650\,{\AA}, with a spectral resolution of 0.3\,{\AA}. For more details, please 
refer to \cite{Napi-etal:01,Napi-etal:03}. 
SPY observations were used to check candidates for our project for spectral 
peculiarities and magnetic fields strong enough to be detected in intensity 
spectra. Hence, our targets were selected with the criterion of not having 
any sign of Zeeman splitting visible in the SPY  spectra, and hence no 
magnetic 
fields above a level of about 20\,kG.

All our targets have strong hydrogen lines, ideal for measuring line 
polarisation, and no peculiarities (such as MG-magnetic fields 
 or a bright companion or any indication of magnetic fields).

\subsection{Data reduction}
Data reduction was performed using {\sc midas} routines. 
Flat field, bias and wavelength calibration exposures were taken during day 
time with the same instrument setup, after each observing night. 

In all frames the bias level was subtracted and  the frames were cleaned 
of cosmic ray hits.
For each observing night a unique master flat-field was calculated from the 
median flux of all flat-fields taken that night. The stellar spectra were 
extracted from the flat-field corrected frames as a sum over about forty 
CCD rows for each beam. Sky spectra were obtained from adjacent regions (about 
twenty CCD rows) on the detector to the observed stars (below and above) and 
subtracted from the object spectra. Wavelength calibration was done using 
HgCd, He and Ar arc spectra, which was independently performed for wavelengths 
of the ordinary and extra-ordinary beams with the corresponding beams of a 
reference spectrum.

Stokes $I$ was obtained as a sum over all beams, while the calculation of 
Stokes ($V/I$) is described in the following section.
\begin{table*}
\begin{center}
\caption{Details of VLT observations. The provided $\alpha$ and $\delta$ 
coordinates refer to epoch 2000 as measured in the course of the SPY project
\citep[see][]{Koester-etal:01}.
Spectral types, $T_{\rm sp}$, and measured V magnitudes were taken from the 
catalogue of \cite{McCook-Sion:99}.} 
\label{t:obs}
\begin{tabular}[c]{llrrrcrrl}
\hline
\hline
\multicolumn{1}{c}{Target} & \multicolumn{1}{c}{Alias} & 
\multicolumn{1}{c}{$\alpha$} & \multicolumn{1}{c}{$\delta$} & 
\multicolumn{1}{c}{V} & \multicolumn{1}{c}{HJD} & 
\multicolumn{1}{c}{$t_{\rm exp}$} & \multicolumn{1}{c}{$n$} & 
\multicolumn{1}{c}{$T_{sp}$} \\
\multicolumn{1}{c}{} & \multicolumn{1}{c}{} & \multicolumn{1}{c}{} &         
\multicolumn{1}{c}{} & \multicolumn{1}{c}{(mag)} & 
\multicolumn{1}{c}{(+2452500)} & 
\multicolumn{1}{c}{(s)} & \multicolumn{1}{c}{} & \multicolumn{1}{c}{} \\
\hline
WD\,0135$-$052 & LHS\,1270     & 01 37 59.4 & $-$04 59 45 & 12.84 & 108.584 & 546 & ~4 & DA7 \\ 
WD\,0227+050   & Feige\,22     & 02 30 16.6 &   +05 15 51 & 12.65 & 137.601 & 381 & ~6 & DA3 \\  
               &               &            &             &       & 169.545 & 381 & ~6 &     \\   
WD\,0310$-$688 & CPD$-$69\,177 & 03 10 31.0 & $-$68 36 03 & 11.40 & 195.536 & 119 & 14 & DA3 \\
WD\,0346$-$011 & GD\,50        & 03 48 50.2 & $-$00 58 31 & 13.99 & 137.657 & 699 & ~4 & DA1 \\
               &               &            &             &       & 174.600 & 699 & ~4 &     \\   
WD\,0446$-$789 & BPM\,03523    & 04 43 46.4 & $-$78 51 50 & 13.47 & 109.710 & 699 & ~4 & DA3 \\
               &               &            &             &       & 168.569 & 699 & ~4 &     \\   
WD\,0612+177   & LTT\,11818    & 06 15 19.0 &   +17 43 48 & 13.39 & 109.756 & 699 & ~4 & DA2 \\
               &               &            &             &       & 172.561 & 699 & ~4 &     \\   
WD\,0631+107   & KPD0631+1043  & 06 33 50.6 &   +10 41 09 & 13.82 & 200.612 & 699 & ~4 & DA2 \\   
               &               &            &             &       & 202.607 & 699 & ~4 &     \\ 
WD\,0839$-$327 & CD$-$32\,5613 & 08 41 32.6 & $-$32 56 35 & 11.90 & 108.803 & 186 & 10 & DA6 \\
WD\,0859$-$039 & RE\,0902-040  & 09 02 17.3 & $-$04 07 12 & 12.40 & 174.709 & 289 & ~8 & DA2 \\   
               &               &            &             &       & 196.701 & 289 & ~8 &     \\ 
WD\,1042$-$690 & BPM\,06502    & 10 44 10.5 & $-$69 18 20 & 13.09 & 168.832 & 426 & ~6 & DA3 \\
               &               &            &             &       & 174.755 & 426 & ~6 &     \\ 
               &               &            &             &       & 195.783 & 426 & ~6 &     \\ 
WD\,1105$-$048 & LP\,672$-$001 & 11 08 00.0 & $-$05 09 26 & 12.92 & 141.838 & 486 & ~4 & DA3 \\
               &               &            &             &       & 169.791 & 486 & ~4 &     \\ 
WD\,2359$-$434 & L\,362$-$81   & 00 02 10.7 & $-$43 09 55 & 13.05 & ~83.510 & 546 & ~4 & DA5 \\
               &               &            &             &       & 108.542 & 546 & ~4 &     \\ 
\hline
\end{tabular}
\end{center}
\end{table*}  
\section{Circular Polarisation}
In order to obtain circular polarisation spectra a Wollaston prism and a 
quarter-wave plate were inserted into the optical path. Each exposure yields 
two spectra on the detector, one from the extra-ordinary beam and the other 
from the ordinary beam.

In the ideal case, the polarisation information (one Stokes parameter per 
exposure) is contained in the ratio, at each wavelength, of the intensities 
in the two spectra (from the ordinary and extra-ordinary beams) but it is 
mixed up with the system gain ratio for the pixels concerned. Alternatively, 
the effect of the not well defined pixel gain can be significantly reduced 
by inverting the sign of the polarisation effects in a second exposure (by 
rotating the $\lambda/4$-plate by $90^\circ$), while leaving the gain ratios 
identical \citep{Tinbergen-Rutten:97}. Hence, Stokes $V$ is obtained from a 
differential measurement of photon counts in either the ordinary or 
extra-ordinary beams, measured at two different angles of the retarder 
waveplate. In this way errors from changes in the sky transparency, 
atmospheric scintillation, and various instrumental effects are significantly 
reduced, so that photon noise remains as the dominant error source.

 Another important source of sytematic error when obtaining Stokes $V$ 
could be the wavelength calibration procedure. An incorrect calibrated 
wavelength scale for each of the two analysed spectra will yield the line 
profiles of the two spectra not to be perfectly aligned, even in the absence 
of magnetic field, leading to spurious polarization signals in every line, 
that change sign as the wave plate rotates. 
No such spurious signals were detected during the calibration process,
confirming that FORS is a very stable instrument. Even if spurious line 
signals at the noise level are present, they would be compensated at least 
partly by the combination of data taken with retarder plate position angles 
$+45^{\circ}$ and $-45^{\circ}$. We note that wavelength calibrations were 
made for each observing date separately. Because most targets were observed 
for two different dates we could check that no spurious magnetic field 
detection due to wavelength calibration errors are present.

We adopted the FORS1 standard observing sequence for circular polarimetry 
consisting of exposures with retarder plate position angles $+45^{\circ}$ and 
$-45^{\circ}$. The number of exposures, $n$, is reported in Table 1 for each 
object. For instance, for $n=4$ the sequence of position angles would go: 
$+45^{\circ}$, $+45^{\circ}$, $-45^{\circ}$, $-45^{\circ}$.  

In order to derive the circular polarisation from a sequence of exposures, we 
added up the exposures with the same quarter-wave plate position angle. 
The Stokes ($V/I$) can be obtained as
\begin{equation}
\frac{V}{I} = \frac{(R-1)}{(R+1)}, ~~\mbox{ with }
R^2=\left(\frac{f_o}{f_e}\right)_{\alpha=+45}\times 
    \left(\frac{f_e}{f_o}\right)_{\alpha=-45}
\label{e:vi}
\end{equation}
where $V$ is the Stokes parameter which describes the net circular 
polarisation, $I$ is the unpolarized intensity, $\alpha$ indicates the nominal 
value of the position angle of the retarder waveplate, and $f_o$ and $f_e$ are 
the fluxes on the detector from the ordinary and extra-ordinary beams of the 
Wollaston, respectively. Eq.\,\ref{e:vi} is equivalent to formula 4.1 in 
the FORS 1+2 User Manual \citep{Szeifert-Boehnhardt:03}. 
 In \cite{Donati-etal:97} a detailed description of the method for the 
suppression of spurious signals in spectropolarimetric observations that we
applied is provided.

The 12 selected white dwarfs are listed in Table 1 together with some 
characteristics of the observations. The Heliocentric Julian Dates correspond
to the beginning of each observing sequence.

\section{Determination of weak magnetic fields}
\subsection{Theoretical polarisation}
For the weak magnetic fields expected in our white dwarfs the longitudinal 
Zeeman effect (due to magnetic fields parallel to the line of sight) is 
measurable for the broad Balmer lines with intermediate spectral resolution. 
This is possible because, in first approximation, the positive and negative 
extrema of Stokes V occur at the wavelengths where the second derivative of 
Stokes I changes its sign. These points could possibly be regarded as the 
transitions between the core and the wings of the line. Hence, the wavelength 
difference between the maxima of the opposite net circular polarisation in the 
two wings of the line, is correlated with the width of the core of the line 
and not with the Zeeman splitting $\Delta \lambda_{\rm B}$, which can be much 
smaller.
The detection threshold therefore depends mainly on the noise level in $V$ 
and the width of the core of the spectral lines, and far less on the achieved 
spectral resolution.

With FORS1 on the VLT we reach a noise level in circular spectropolarimetry 
of about $3\times 10^{-3} I_{\rm c}$, where $I_{\rm c}$ is the continuum 
intensity. This allows us to detect magnetic fields of a few kG for the 
brightest white dwarfs in our sample with exposure times of about 1 hour. 
This is a significant improvement compared to the observations of 
\cite{Schmidt-Smith:95}, who reached a detection limit of about 20 kG 
($2.3\sigma$ limit) with a 2~m class telescope. By adding the signal of 
several Balmer lines, which are observed simultaneously with our instrument 
setup, the magnetic sensitivity can be further enhanced. However, the error 
of the magnetic field determination increases for the higher series numbers, 
so that a reliable analysis can be based on H$\beta$ and H$\gamma$ only.

The theory of spectral line formation in a magnetic atmosphere shows that the 
splitting of a spectral line observed in both senses of circular polarisation 
is proportional to $\langle B_z \rangle$, the average of the component of the 
magnetic field along the line of sight averaged over the visible stellar 
hemisphere, \ie, the mean longitudinal magnetic field 
\citep{Babcock:47,Babcock:58}. 

For field strengths below 10\,kG the Zeeman splitting of the Balmer lines is 
less than approximately 0.1 \ang. This is well below the width of the cores 
of the Balmer lines in all the stars of our sample (typically a few \ang). 
Therefore, we can apply the weak-field approximation 
\citep[\eg,][]{Angel-Landstreet:70,Landi-Landi:73} without any loss of 
accuracy. According to this approximation the measured $V$ and $I$ profiles 
are related to $\langle B_z\rangle$ by the expression:
\begin{equation}
\frac{V}{I} = -g_{\rm eff} \ensuremath{C_z}\lambda^{2}\frac{1}{I}
\frac{\partial I}{\partial \lambda}
\ensuremath{\langle\large B_z\large\rangle}\;,
\label{e:lf}
\end{equation}
where $g_{\rm eff}$ is the effective Land\'{e} factor 
\citep[= 1 for all hydrogen lines of any series,][]{Casini-Landi:94}, 
$\lambda$ is the wavelength expressed in \ang, $\langle B_z\rangle$ is the 
mean longitudinal component of the magnetic field  expressed in Gauss and the 
constant 
$C_z=e/(4\pi m_ec^2)$ $(\simeq 4.67 \times10^{-13}\,{\rm G}^{-1} \ang^{-1})$.
Note that this approximation also holds if instrumental broadening is present, 
but it is not generally correct if the profiles are rotationally broadened 
\citep{Landstreet:82}.

Eq.\,\ref{e:lf} is evaluated by using the high signal-to-noise spectra 
$I(\lambda)$ and their derivative $\partial I/\partial \lambda$. For a given 
wavelength $\lambda_i$ we approximate $\partial I/\partial \lambda$ by the 
average of $(I_{i+1}-I_i)/(\lambda_{i+1}-\lambda_i)$ and 
$(I_i-I_{i-1})/(\lambda_i-\lambda_{i-1})$.

The error associated with the determination of the longitudinal field obtained 
from individual Balmer lines is larger for Balmer lines at shorter wavelengths 
than for lines at longer wavelengths. This is due to the combination of two 
effects: while the Zeeman  effect increases as lambda squared, most other 
line 
broadening effects depend linearly on lambda, so that the magnetic field is 
better detected at longer wavelengths than at shorter wavelengths; 
furthermore, the Balmer lines at shorter wavelengths are less deep, so that 
$\partial I/\partial \lambda$ is smaller. Using H$\beta$ and H$\gamma$ 
simultaneously, we obtained a determination of the mean longitudinal magnetic 
field that best fit the observed ($V/I$) (as explained in the next section).

\begin{figure*}
\epsfxsize=16.cm \epsfbox{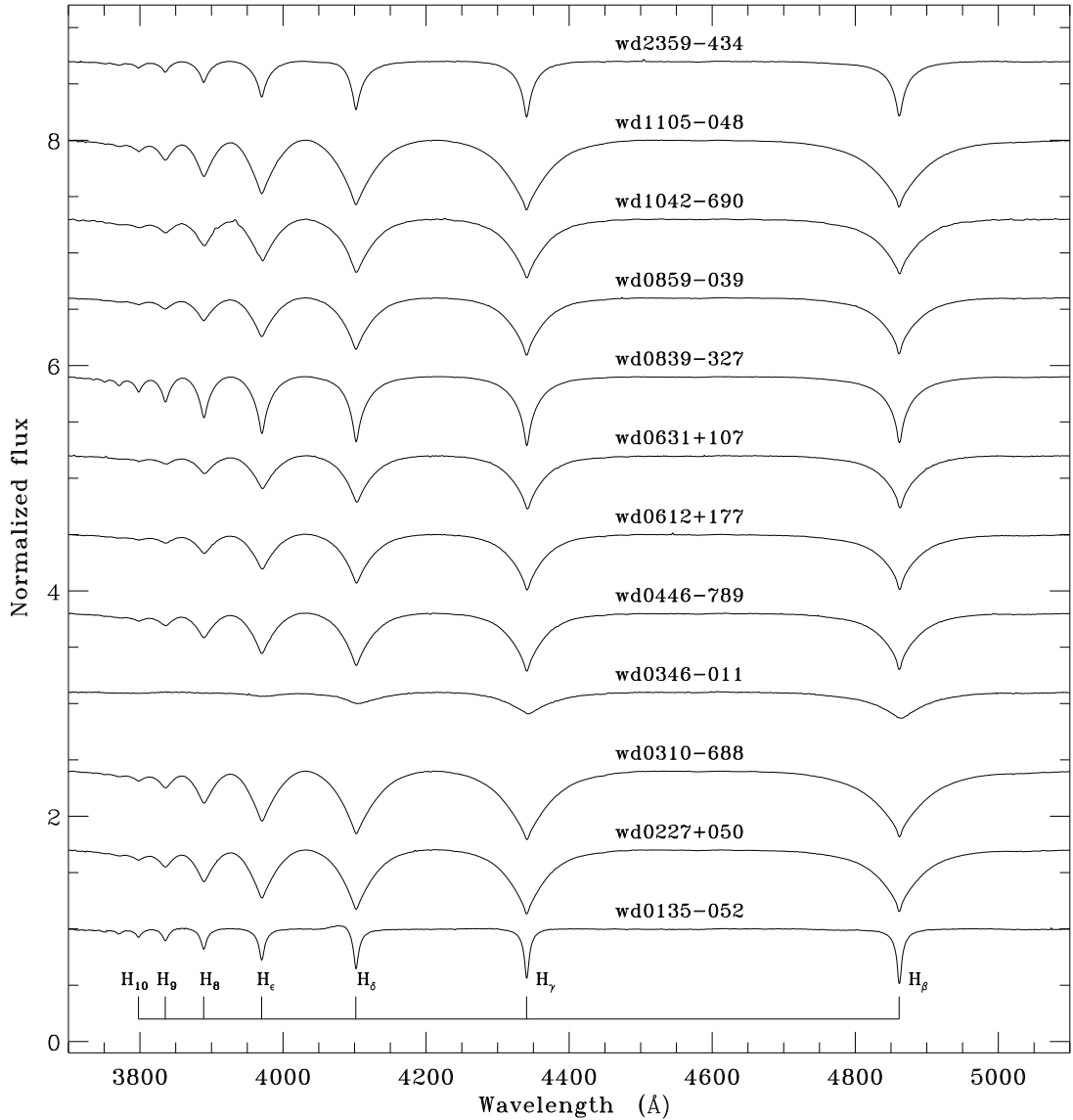}
\caption{Normalised spectra of our sample of white dwarfs in the region of 
the Balmer series. When more than one sequence of exposures was available for
a white dwarf, the average spectrum is presented. The positions of the Balmer 
lines are indicated. The spectra are displaced vertically by 0.7 units (or 
multiples thereof) for a better visualisation.}
\label{f:ni}
\end{figure*}
\subsection{Fitting of the magnetic field}
In Fig.\,\ref{f:ni} we present the normalised spectra of our sample of 12 
DA white dwarfs. Multiple observations of the same star were added up 
in order to achieve a higher signal-to-noise level. In some spectra,
particularly prominent in WD 0135-052, a small bump is visible on the 
blue side of the H$\delta$ line, due to the reflex mentioned in Sect. 2.

For the circular polarisation spectra ($V/I$) flux weighted means were
calculated according to the following formula for two measurements 1
and 2:
\begin{equation}
\ensuremath{\left(\frac{V}{I}\right)_{\rm tot}} = 
\frac{\left(\frac{V}{I}\right)_{1}~I_{1}+\left(\frac{V}{I}\right)_{2}~I_{2}}
     {I_{1}+I_{2}}.
\label{e:av}  
\end{equation}
For the averaging we have checked that the effect of the different barycentric 
corrections for multiple observations of the same object is much smaller than 
our spectral resolution.
The averaging helps to extract signal hidden in the noise of the individual 
exposures if the stellar Zeeman signal remains unchanged with time.

\begin{figure*}
\epsfxsize=16.cm \epsfbox{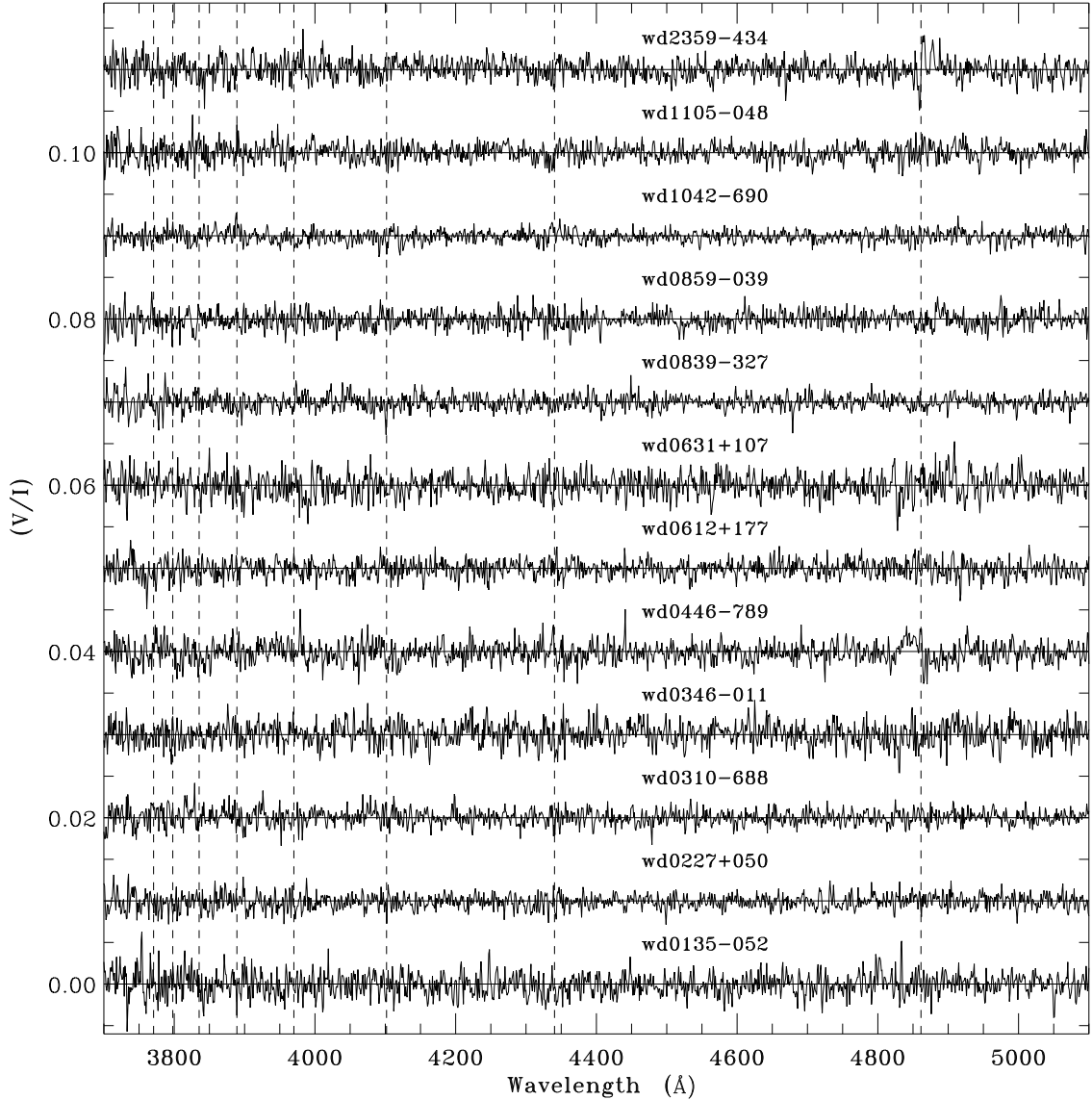}
\caption{Circular polarisation spectra ($V/I$) of our sample of white dwarfs 
in the region of the Balmer series. The average ($V/I$)-spectrum is plotted 
for multiple observations. For all spectra a horizontal line is drawn to 
indicate the zero level. Dashed vertical lines represent the positions where 
the Balmer lines are located. Spectra are displaced with relative shifts for 
a better visualisation.}
\label{f:vi}
\end{figure*}

Fig.\,\ref{f:vi} shows the circular polarisation spectra ($V/I$) of our sample 
of white dwarfs in the region of the Balmer series (no binning along the 
spectrum was applied to 
the data). The positions of the Balmer lines are indicated by vertical dashed 
lines. In two of the averaged spectra (corresponding to stars WD\,0446$-$789 
and WD\,2359$-$434) a moderate S-shape circular polarisation signature across 
H$\beta$ and H$\gamma$ can be noticed. For WD\,1105$-$048 polarisation 
reversals of those lines were only present in the observation of 9 Jan. 2003 
(see Fig.\,\ref{f:B_wd1105}).

In order to determine the longitudinal component of the magnetic field  for
each measurement we compared the observed circular polarisation for an 
interval of $\pm 20$\,\ang\ around H$\beta$ and H$\gamma$ with the prediction 
of Eq.\,\ref{e:lf}. The best fit for $\langle B_z\rangle$, the only free 
parameter, was found by a $\chi^2$-minimisation procedure. 
If we assume that no magnetic field is present, all deviations from zero 
polarization are due to noise. This can be expressed by the standard deviation 
$\sigma$ over the respective intervals around the Balmer lines. If the best 
fit magnetic field is indeed close to zero, the reduced $\chi^2$ should be 
automatically close to unity. However, in the case of a finite magnetic field 
not all deviations from zero would be due to noise, so that the best fit would 
have a reduced $\chi^2$ smaller than 1.
Following \cite{Press:86} we determined the statistical error from the rms 
deviation of the observed circular polarisation from the best-fit model.
The $1\sigma$ (68.3\%) confidence range for a degree of freedom of 1 is the 
interval of $B_z$ where the deviation from the minimum is $\Delta \chi^2=1$. 
Note that this is a purely statistical error and does not account for
systematic errors, e.g. from the limitation of our low-field approximation
which, however, we expect to be  rather small, or from wavelength 
calibration errors.
  
The results for each of our single observations is summarized in 
Table\,\ref{t:mf}, where we have listed the best fits for $B_z$ for the 
H$\beta$ and H$\gamma$ lines as well as the weighted value 
$B_z=(B_{z,\gamma} w_\gamma+B_{z,\beta} w_\beta)/(w_\gamma+w_\beta)$ where 
$w_i=1/\sigma_i^2$ ($i=\gamma,\beta$). The probable error is given by 
$\sigma=(\sum w_i)^{-1/2}$. When multiple observations were added up to reduce 
the noise level (by means of Eq.\,\ref{e:av}), results are labeled as
{\sc average}. 
\begin{table*}
\begin{center}
\caption{Magnetic fields derived from the H$\gamma$ and H$\beta$ lines for our 
sample of white dwarfs. $B({\sigma})$ provides the magnetic field in units of 
the $\sigma$ level. Detections exceeding the $3\sigma$ levels are given in 
bold. $\sigma(V)$ is the standard deviation of the observed ($V/I$)-spectrum 
obtained in the region 4500--4700 \ang. Lower limits on the detectability of 
the magnetic field from the line polarisation peaks, calculated at the 
$1\sigma$ level of the noise, are given in the last two columns. Multiple 
observations that were averaged prior to analysis are labeled {\sc average}.}
\label{t:mf}
\begin{tabular}[c]{lccrrrrrr}
\hline
\hline
\multicolumn{1}{c}{Target} & \multicolumn{1}{c}{Date} & 
\multicolumn{1}{c}{$\sigma(V)$} & \multicolumn{2}{c}{$B$(G)} & 
\multicolumn{1}{c}{$B({\rm G})$} & \multicolumn{1}{c}{$B({\sigma})$} &
\multicolumn{2}{c}{B(G) at $1\sigma$}\\
\multicolumn{1}{c}{} & \multicolumn{1}{c}{} & 
\multicolumn{1}{c}{(10$^{-3}I_{\rm c}$)} & 
\multicolumn{1}{c}{H$\gamma$} & \multicolumn{1}{c}{H$\beta$} & 
\multicolumn{1}{c}{H$\gamma, \beta$} & \multicolumn{1}{c}{H$\gamma, \beta$} & 
\multicolumn{1}{c}{~~~H$\gamma$} & \multicolumn{1}{c}{~~~H$\beta$} \\
\hline
WD\,0135$-$052 & 29/11/02 & 1.2 &   $1112\pm625$ &    $492\pm373$ &    $654\pm320$ & 2.04 & 1277 &   861 \\[2pt]
WD\,0227+050   & 28/12/02 & 1.0 &   $493\pm1177$ &   $-776\pm707$ &   $-439\pm606$ & 0.72 & 2486 &  $-$1928 \\
               & 29/01/03 & 1.0 &   $-710\pm971$ &   $1854\pm709$ &    $962\pm572$ & 1.68 & $-$2969 &  1989 \\
          & {\sc average} & 0.7 &   $-105\pm789$ &    $590\pm469$ &    $408\pm403$ & 1.01 & $-$1924 &  1194 \\[0.08cm]
WD\,0310$-$688 & 24/02/03 & 0.8 &   $-504\pm741$ &    $114\pm548$ &   $-104\pm440$ & 0.24 & $-$2106 &  1404 \\[0.08cm]
WD\,0346$-$011 & 28/12/02 & 1.8 & $-9299\pm8510$ & $-6815\pm5295$ & $-7508\pm4495$ & 1.67 & $-$36860 & $-$16600 \\
               & 03/02/03 & 2.6 &  $3222\pm9374$ &  $7542\pm6717$ &  $6076\pm5459$ & 1.11 & 28440 & 24820 \\
          & {\sc average} & 1.1 & $-5967\pm7087$ & $-1496\pm4226$ & $-2668\pm3629$ & 0.74 & $-$23060 & $-$10680 \\[0.08cm]
WD\,0446$-$789 & 30/11/02 & 1.3 &   $417\pm1403$ &   $4251\pm910$ &   {\bf 3115$\pm$763} & 4.08 & 3731 &  2340 \\
               & 28/01/03 & 1.4 &  $7771\pm1476$ &  $5370\pm1196$ &   {\bf 6321$\pm$929} & 6.80 & 4118 &  2815 \\
          & {\sc average} & 0.9 &  $3612\pm1005$ &   $4743\pm832$ &   {\bf 4283$\pm$640} & 6.69 & 2495 &  1594 \\[0.08cm]
WD\,0612+177   & 30/11/02 & 1.3 &  $2416\pm1456$ & $-1148\pm1203$ &    $297\pm927$ & 0.32 & 4060 &  $-$2620 \\
               & 01/02/03 & 1.4 &    $86\pm1154$ &   $236\pm1171$ &    $156\pm821$ & 0.19 & 3742 &  2915 \\
          & {\sc average} & 0.9 &   $1132\pm920$ &   $-457\pm789$ &    $216\pm598$ & 0.36 & 2672 &  $-$1939 \\[0.08cm]
WD\,0631+107   & 01/03/03 & 2.2 & $-1310\pm1785$ &  $-170\pm1815$ &  $-749\pm1272$ & 0.59 & $-$7300 &  $-$5890 \\
               & 03/03/03 & 2.1 &  $-790\pm1870$ &   $310\pm1775$ &  $-211\pm1287$ & 0.16 & $-$6630 &  4340 \\
          & {\sc average} & 1.2 &  $-990\pm1440$ &   $260\pm1090$ &   $-195\pm869$ & 0.22 & $-$4010 &  2630 \\[0.08cm]
WD\,0839$-$327 & 29/11/02 & 0.7 &   $-313\pm325$ &    $402\pm261$ &    $121\pm203$ & 0.60 & $-$1008 &   761 \\[0.08cm]
WD\,0859$-$039 & 03/02/03 & 1.5 &  $2083\pm1164$ &     $11\pm987$ &    $877\pm752$ & 1.17 & 4400 &  2727 \\
               & 25/02/03 & 1.2 &  $2183\pm1372$ &   $-401\pm911$ &    $389\pm758$ & 0.51 & 3234 &  $-$2222 \\
          & {\sc average} & 0.8 &   $1186\pm928$ &    $382\pm607$ &    $622\pm507$ & 1.23 & 2319 &  1511 \\[0.08cm]
WD\,1042$-$690 & 28/01/03 & 1.3 &  $-719\pm2333$ & $-2850\pm2301$ & $-1799\pm1638$ & 1.10 & $-$3624 &  $-$3247 \\
               & 03/02/03 & 1.0 &  $1627\pm1464$ &  $-865\pm1163$ &     $99\pm910$ & 0.11 & 2963 &  $-$3560 \\
               & 24/02/03 & 0.9 & $-2374\pm1060$ &    $103\pm864$ &   $-885\pm669$ & 1.32 & $-$2764 &  1804 \\
          & {\sc average} & 0.6 &    $141\pm745$ &   $-406\pm620$ &   $-182\pm476$ & 0.38 & 1833 &  $-$1508 \\[0.08cm]
WD\,1105$-$048 & 01/01/03 & 1.1 &  $-1153\pm632$ &     $55\pm807$ &   $-693\pm497$ & 1.39 & $-$2863 &  1599 \\
               & 29/01/03 & 1.1 & $-4686\pm1032$ &  $-3305\pm979$ &  {\bf $-$3959$\pm$710} & 5.58 & $-$2756 &  $-$2164 \\
          & {\sc average} & 0.8 &  $-2760\pm598$ & $-1337\pm675$  &  {\bf $-$2134$\pm$447} & 4.77 & $-$2156 &  $-$1538 \\[0.08cm]
WD\,2359$-$434 & 04/11/02 & 1.0 & $-3003\pm1433$ & $-5719\pm1289$ &  {\bf $-$4504$\pm$958} & 4.70 & $-$2914 &  $-$2975 \\
               & 29/11/02 & 1.0 &  $-1905\pm702$ &  $-4224\pm796$ &  {\bf $-$2919$\pm$526} & 5.55 & $-$4880 &  $-$3907 \\
          & {\sc average} & 1.0 &  $-2066\pm546$ &  $-4738\pm667$ &  {\bf $-$3138$\pm$422} & 7.44 & $-$1785 &  $-$1532 \\ 
\hline
\end{tabular}
\end{center}
\end{table*}  

\begin{figure}[!h]
\epsfxsize=9.cm \epsfbox{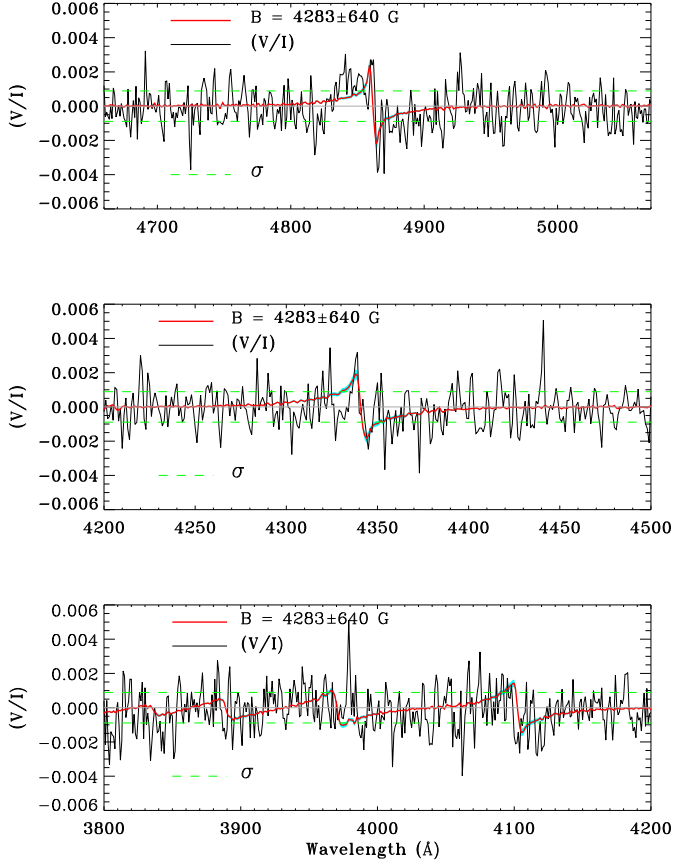}
\caption{Circular polarisation spectra ($V/I$) of  WD\,0446$-$789 (average 
of observations from 30/11/03 and 28/01/03, thin solid line) in the region 
of H$\beta$ (top), H$\gamma$ (middle) and close to the Balmer jump (bottom 
panel). The solid horizontal line indicates the zero level. The horizonal 
dashed lines indicate the position of the $1\sigma$ level of the average 
($V/I$) spectrum. The light shading (cyan in the electronic version) 
represents the variation between ($V/I$) spectra predicted by the low-field 
approximation (Eq.\,\ref{e:lf}) using magnetic field values of 4283$\pm$640\,G 
(thick solid line, red in the electronic version).}
\label{f:wd0446}
\end{figure}

\begin{figure}[!h]
\epsfxsize=9.cm \epsfbox{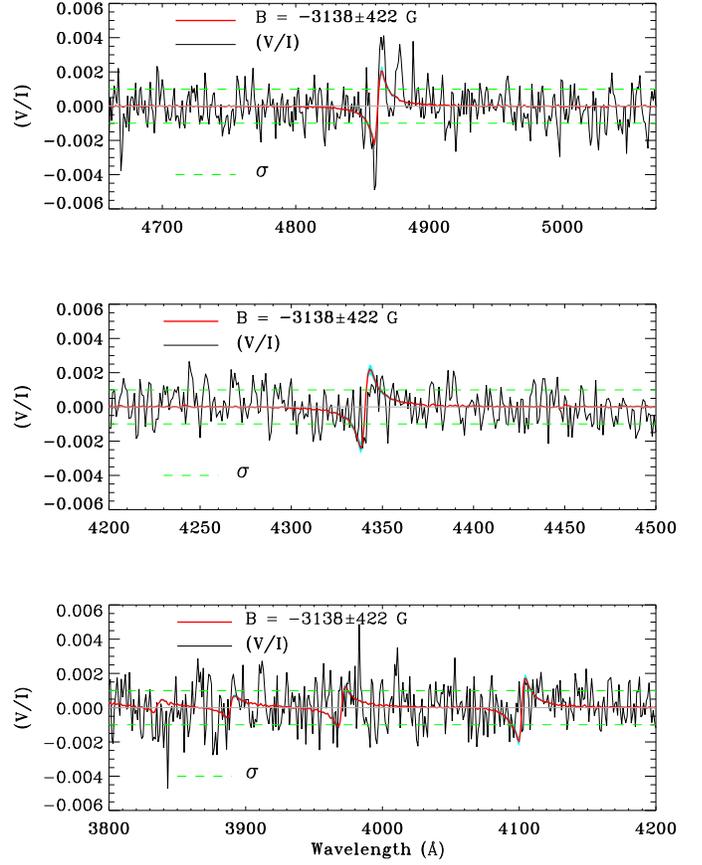}
\caption{As in Fig.\,\ref{f:wd0446} but for WD\,2359$-$434 (average of 
observations from 04/11/03 and 29/11/03), where the best fit is at 
$\langle B_z\rangle=-3138\pm422$\,G.} 
\label{f:B_wd2359}
\end{figure}

\begin{figure}[!h]
\epsfxsize=9.cm \epsfbox{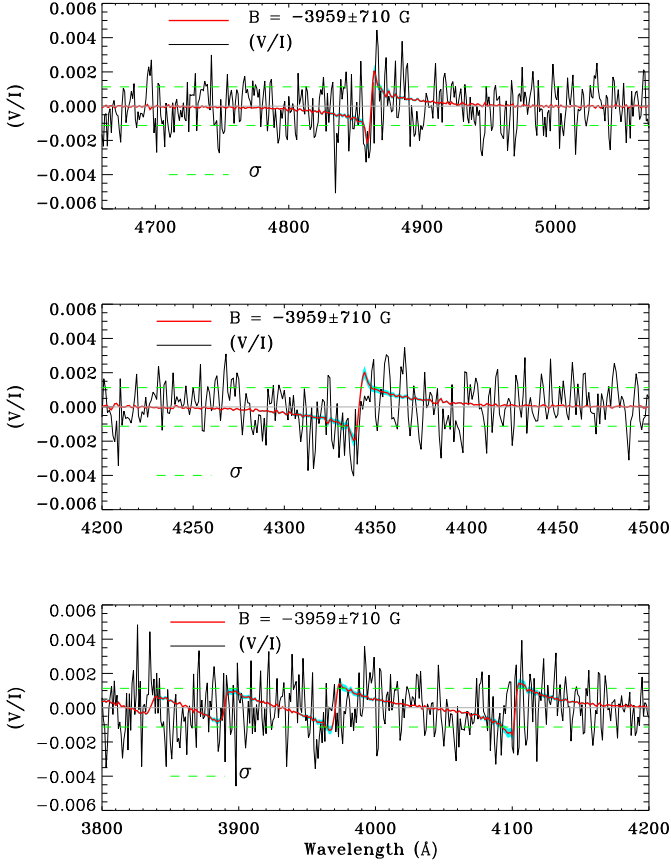}
\caption{As in Fig.\,\ref{f:wd0446} but for the 09/01/03 observation of the 
white dwarf WD\,1105$-$048 where the best fit is at 
$\langle B_z\rangle=-3959\pm710$\,G.} 
\label{f:B_wd1105} 
\end{figure}

In three stars we find a significant magnetic field: WD\,0446$-$789 
($4283\pm640$\,G, see Fig.\,\ref{f:wd0446}), WD\,2359$-$434 ($-3138\pm422$\,G, 
see Fig.\,\ref{f:B_wd2359}), and WD\,1105$-$048 ($-2134\pm447$\,G). 
Positive and negative signs of $B_z$ indicate opposite magnetic polarity. 
For the first two stars the magnetic field is detected at the $3\sigma$ level 
individually from H$\beta$ and H$\gamma$ (at least for the averaged spectra), 
as well as from the combination of both lines. This increases our confidence 
in both detections (cf. Sect. 4.3). 

Although several lines of the Balmer series are available, the higher 
members of the series do not contain enough $V$-signal to give reliable 
results. However, the analysis of H$\delta$ (the only line among the higher 
members with its $V$-signal exceeding the $1\sigma(V)$), confirms the positive 
detections of magnetic fields in the three objects WD\,0446$-$789, 
WD\,1105$-$048 and WD\,2359$-$434.

In WD\,0446$-$789 the analysis of the 30 Nov. 2002 observation around 
H$\gamma$ yields a magnetic field of $417\pm1403$\,G, corresponding to a 
$\chi^2_{\rm min}=1.0$ (instead of $\chi^2_{\rm min}=0.6$ for 
the 28 Jan. 2003 observation), \ie\ a $0.1\sigma$ level detection. If we do 
not use this outlier for our combined result from the two observations, we 
obtain a $3\sigma$ detection of $4743\pm832$\,G.

In the case of WD\,1105$-$048 the different values of the magnetic field 
obtained for the two observations on 1 Jan. 2003 ($-693\pm497$\,G) and 29 Jan. 
2003 ($-3959\pm710$\,G, see Fig.\,\ref{f:B_wd1105}) differ significantly. This 
may indicate a different orientation of the magnetic field between the two 
epochs due to stellar rotation.

The case of WD\,0135$-$052, where the formal value for the combined result 
from H$\beta$ and H$\gamma$ exceeds the $1\sigma$ error by a factor of two, is 
rather uncertain since it is based on one observation only. In addition, this 
object is a double-lined binary with a period of 1.556 days and it is not 
clear whether the orbital variation can mimic a magnetic field.
 We agree with the referee, who pointed out that this could just be a 
$2\sigma$ level detection, which can be expected for our sample size of 
field measurements just by chance. However, 
WD\,0135$-$052 is a good candidate for follow-up observations, although one 
must bear in mind that the magnetic field would be diluted by the presence of 
a non-magnetic companion. 

In all other stars in our sample the best-fit magnetic field is below or close 
to the $1\sigma$ error.

In Figs.\,\ref{f:wd0446} to \ref{f:art} the observed circular polarisation 
spectra of the white dwarfs (thin solid line) are presented together with the 
($V/I$) spectra predicted by the low-field approximation (thick solid line, 
red colour in the electronic version) for the given value of $B$. The light 
shading (cyan in the electronic version) shows the values of ($V/I$) obtained 
varying $B$ by the indicated errors, and can be more clearly distinguished 
from the predicted ($V/I$) spectra at the peaks of the S-wave features 
(in particular in the region of H$\gamma$).
 

\subsection{Accuracy of the fitting procedure}
Although the $\chi^2$ fitting provides formal errors, it is difficult to judge 
whether this error actually provides a reasonable estimate of the confidence 
range for our purpose. For example, the choice of our interval of 
$\pm$20\,\ang\ around H$\beta$ and H$\gamma$ is arbitrary. Therefore, we 
performed simulations for different magnetic fields, interval sizes, and noise 
levels in order to test our approach. The Gaussian noise for our simulated 
polarisation spectra was provided by the random number  generator implemented 
in IDL. 

As an example we used the solution for the average of  both WD\,0446$-$789 
observations (B=4283$\pm$640 G), its noise level ($\sigma=0.001$ at both 
H$\beta$ and H$\gamma$) and calculated 1000 simulated polarisation spectra 
for the interval size of $\pm$20\,\ang. The average result for the magnetic 
field was: 4301$\pm$442 G (solution for H$\gamma, \beta$), 4276$\pm$554 G 
(for H$\beta$ only), and 4339$\pm$725\,G (for H$\gamma$ only). The mean value 
of B agrees very well with the prescribed value. The mean of the 
individual standard deviations of the 1000 runs are 413\,G, 541\,G and 667\,G, 
for H$\gamma, \beta$, H$\beta$ and H$\gamma$, respectively. This is slightly 
smaller than 640\,G, 832\,G and 1005\,G from the WD\,0446$-$789 observation. 
This may be due to the fact that we assume a homogeneous magnetic field which 
may not be the case in reality, or it may indicate a small hidden source of 
errors in the data (\eg\ a non-Gaussian distribution of the noise).

\begin{figure}[!h]
\epsfxsize=9.cm \epsfbox{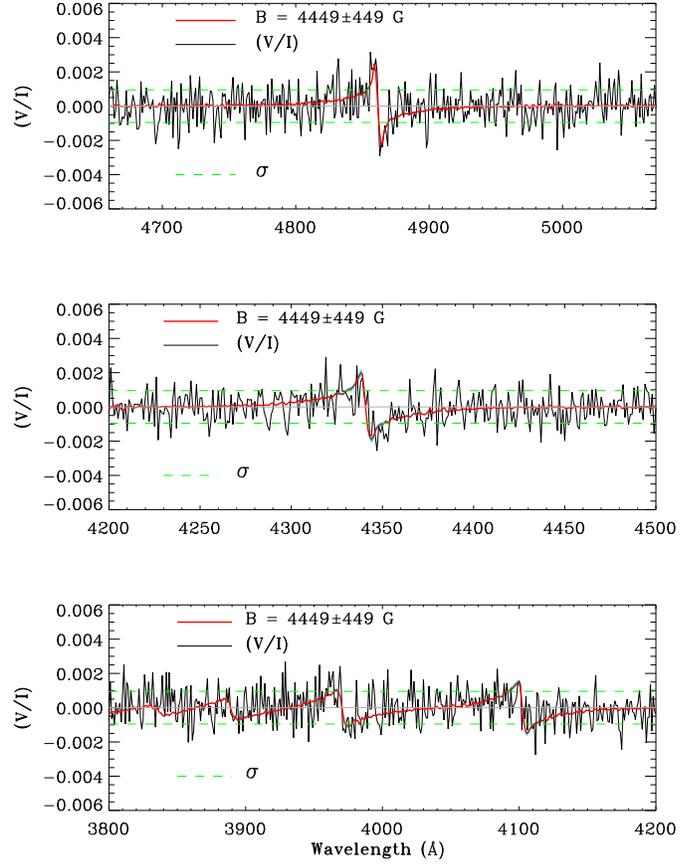}
\caption[]{Artificial polarisation spectrum with the same noise level as 
WD\,0446-789 plotted in Fig.\,\ref{f:wd0446}. For this particular example, 
out of the 1000 simulations, the fitting procedure results in 
$\langle B_z\rangle=4449\pm449$\,G.}
\label{f:art}
\end{figure}

Fig.\,\ref{f:art} shows one of the artificial polarisation measurements from 
our 1000 simulations. The fit is 4449$\pm$449 G in this particular example. 
It demonstrates that a highly significant detection is possible at our noise 
level.

The resulting magnetic field changed by less than 2\%\ if an interval of $\pm$ 
40 \ang\ around H$\beta$ and H$\gamma$ was used. In the case of a 
smaller interval, however, larger deviations occur since not the whole range 
where the predicted ($V/I$) differs from zero is included.

We are aware of the fact that the polarisation feature is mostly buried in 
noise in the ($V/I$)-spectrum, with the exception of the narrow peaks on both 
sides of the hydrogen line centers. Therefore, on a plot the magnetic 
information is, to a large extent, invisible  to the eye since in the 
wings 
the solution is based on an average small excess of right- and left-handed 
polarisation on different side of the line core, respectively, which clearly 
contributes to our $\chi^2$ analysis. 
However, if one wants to obtain an idea 
of what magnetic field is needed so that the predicted polarisation peaks 
exceed the $1\sigma$ level of the noise, we provide lower limits for this 
detectability in columns 8 and 9 of Table\,\ref{t:mf}. 
The $\sigma$ was obtained as the standard deviation of the observed 
($V/I$)-spectrum in a region with no lines over 200 {\AA}\ (\ie\ 4500-4700 
\ang), for each single observation.

We note that in the three objects (WD\,0446$-$789, WD\,1105$-$048 and 
WD\,2359$-$434) with a positive detection according to the $\chi^2$ analysis, 
the peaks of the S-wave circular polarisation signature reaches the $2\sigma$ 
level of the noise so the detection is, therefore, confirmed by this very
conservative approach. For the averaged spectra of WD\,0446$-$789 and 
WD\,2359$-$434, the detections derived from the more reliable H$\beta$ line 
even reach the $3\sigma$ level.

\section{Atmospheric and stellar parameters for the target stars}
For a discussion of the evolutionary status of our white dwarfs with and 
without detected magnetic fields we need to know the fundamental parameters 
of our objects. Masses and cooling ages are of special interest to distinguish 
between the proposed formation scenarios. These quantities can be computed 
from the fundamental stellar parameters temperature and gravity, which can be 
derived by a model atmosphere analysis of the spectra, and theoretical cooling 
tracks.

Since all white dwarfs selected for our project are bright by white dwarf 
standards, at least one model atmosphere analysis was published for each 
object. This is a somewhat inhomogeneous collection of data relying on 
spectra of different quality and a variaty of methods, including analysis of 
UV spectra and parallax measurements. Most of the programme stars have already 
been observed with the SPY project. Although our spectra have lower resolution 
than the SPY spectra, which aim at the measurement of radial velocity 
variations seen in the NLTE core of H$\alpha$, our flux spectra have a very 
high signal-to-noise ratio, allowing a very accurate determination of the 
effective temperatures and gravities.

The observed line profiles are fitted with theoretical spectra from a 
large grid of NLTE spectra calculated with the NLTE code developed by 
\cite{Werner:86}. Basic assumptions are those of static, plane-parallel 
atmospheres in hydrostatic and radiative equilibrium. The adopted chemical 
composition is pure hydrogen, which is appropriate for the stars of our 
sample. A description of the model calculations and the adopted atomic physics 
is given in \cite{Napi-etal:99}, where a discussion of the impact of NLTE on 
the atmospheres and line profiles of DA white dwarfs is provided as well. The 
coolest four white dwarfs of our sample  (WD\,0310$-$688, WD\,0839$-$327,
WD\,1105$-$048 and WD\,2359$-$434) were analysed with a grid of LTE model 
spectra computed by D.~Koester for the analysis of DA white dwarfs. The input 
physics of these model atmospheres is described in some detail in 
\cite{Finley-etal:97}. LTE models are more reliable below 17000\,K, because 
the NLTE models do not take into account convection and collision induced
absorption by hydrogen quasi-molecules. On the other side, NLTE effects are 
small at these temperatures (Napiwotzki et al.\ 1999) and of no relevance for 
our results.

The line fits were performed with a modified version of the least-squares 
algorithm of \cite{Bergeron-etal:92} described in \cite{Napi-etal:99}. The 
observed and theoretical Balmer line profiles are normalised to a linear 
continuum in a consistent manner. Radial velocity offsets are corrected by 
shifting the spectra to a common wavelength scale. The synthetic spectra are 
convolved to the resolution of the observed spectra (4.5\,{\AA}) with a Gaussian 
and interpolated to the actual parameters. The atmospheric parameters are then 
determined by minimising the $\chi^2$ value by means of a Levenberg-Marquardt 
steepest descent algorithm \citep{Press:86}. 
 This procedure is applied simultaneously to all Balmer lines of one
observed spectrum. Formal errors can be derived from 
the covariance matrix. However, due to the very high signal-to-noise ratios of 
our spectra, the estimated statistical errors are extremely small (not larger 
than 20\,K in temperature and 0.005\,dex in the log of gravity) and do not 
provide realistic estimates for the whole error budget. The real external 
errors for the white dwarfs investigated in this article can be estimated to 
be 2.3\% in \Teff\ and 0.07\,dex in $\log g$ \citep{Napi-etal:99}. 

\begin{figure}
\epsfxsize=8.5cm \epsfbox{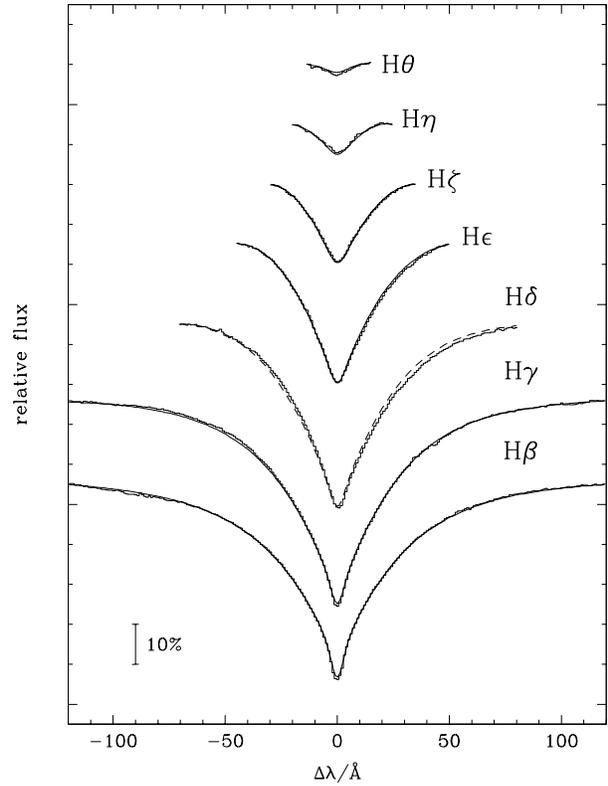}
\caption[]{Model atmosphere fit (solid line) of the 30/11/02 observation of 
WD\,0446$-$789 (solid histogram). The dashed model profile for H$\delta$ 
indicates that this line was not used during the fitting procedure.}
\label{wd:0446fit}
\end{figure}

Our fit results are provided in columns two and three of Table~\ref{t:lit}. 
The parameters are the straight average of the fit results for the 
individual spectra. For the 
reasons discussed above, we do not give the formal fit errors. WD\,0135$-$052 
was not fitted by us, because this is a well known double-lined binary 
consisting of two DA stars \citep{Saffer-etal:88}, which requires a special
treatment. White dwarf masses were computed from a comparison of parameters 
derived from the fit with the grid of white dwarf cooling sequences of 
\citet{Benvenuto+Althaus:99} for an envelope hydrogen mass of 
$10^{-4}M_{\mathrm{WD}}$, and are given in column four of Table~\ref{t:lit}.  
A sample fit is shown in Fig.~\ref{wd:0446fit}. Most Balmer lines are fitted 
very well, but the observed profile of H$\delta$ is only poorly reproduced. 
This results from the reflex in the FORS1 optics mentioned in Sect. 2. Thus we
excluded this line from our fitting procedure. However, parameter changes are 
small, if H$\delta$ is included anyway. 
Spectroscopic  distances $d$(spec) are determined 
from the absolute magnitudes computed for the given stellar parameters using 
the synthetic $V$ band fluxes computed by \citet{Bergeron-etal:95} and the 
measured $V$ magnitudes given in our Table~1, and are provided in column five 
of Table~\ref{t:lit}. Note that the error limits for \Teff\ and $\log g$ given 
above correspond to a 7\% error in the distance.

Table\,\ref{t:lit} is supplemented by physical parameters of our sample of 
white dwarfs collected from the literature. Fundamental parameters such as 
temperature and gravity determined by previous spectroscopic studies are given 
in columns six and seven. The agreement is generally good. Differences are 
usually within the error limits discussed above. Rotational velocities \vsini, 
if available, are shown in column eight. If parallax measurements exist, the 
resulting distances  d(trig) are given in column nine (H indicates 
Hipparcos values, while Y are ground based data from the Yale general 
catalogue of trigonometric parallaxes).

\begin{table*}
\begin{center}
\caption{Fitted parameters of the white dwarfs and supplementary data from 
literature.  The three objects with positive detections of magnetic fields 
are labelled in bold face.} 
\label{t:lit}
\begin{tabular}[c]{l|rlcrr|r@{.}ll|cc}
\hline
\hline
   && &&&&\multicolumn{5}{c}{literature} \\
WD & T$_{\rm eff}$ & \logg & $M$ & $d$(spec) &$t_{\mathrm{cool}}$
   & \multicolumn{2}{c}{T$_{\rm eff}$} 
   &\logg & \vsini & $d$(trig)\\
   &(k\kelvin) && (\Msolar) &(pc) &(Myr)
   &\multicolumn{2}{c}{(k\kelvin)} & & (\kmsec) &(pc)\\
\hline
0135$-$052 && & 0.47/0.52$^{1,14}$ & &
         &\multicolumn{2}{c}{7.47/6.92}  
         &7.80/7.89$^{1,14}$ & & $12.3\pm0.4$Y$^{15}$\\
0227+050 &18.45& 7.79 &0.51&26.1 &65
         &19&07 &7.78$^{2}$   & & $24.3\pm 2.9$H$^{16}$ \\
    &&&&&&18&21  &7.95$^{16}$  &&\\
0310$-$688 &15.73&8.00 &0.61 &11.2 &170
         & 15&71 & 8.16$^{16}$ & 0$\pm$4$^{10}$ & $10.2\pm 0.2$H$^{16}$ \\
    &&&&&& 16&18 & 8.06$^{4}$ &  & \\
0346$-$011 &43.17& 9.08 &1.24 &34.0 &59
         & 39&51 & 9.07$^{12}$ & $\leq$50$^{17}$ & \\
    &&&&&& 43&20 &9.21$^{18}$ & & \\
    &&&&&& 40&54 & 9.22$^{2}$ &  &  \\
    &&&&&& 43&10 & 9.09$^{4}$ & &  \\
{\bf 0446$-$789} &23.45&7.72 & 0.49 &48.8 &21
         & 23&20 & 7.70$^{17}$  & &  \\
    &&&&&& 23&61 & 7.83$^{4}$ & & \\
0612+177 &25.58&7.86 & 0.56 &46.1 &17
         & 24&66 & 7.94$^{2}$  & & $36.1\pm 2.9$Y$^{15}$ \\
    &&&&&& 25&17 & 7.83$^{5}$ & &  \\
    &&&&&& 25&94 & 7.97$^{4}$ & & \\
0631+107 &27.25&7.81 & 0.54 &61.9 &13
         & 26&59 & 7.82$^{5}$ & &  \\
    &&&&&& 27&20 & 8.00$^{8}$   & &  \\
0839$-$327 &9.24& 7.90 & 0.54 &8.3 &660
         & 8&80$^{7}$          & & & $8.9\pm 0.8$Y$^{15}$ \\
    &&&&&& 8&93 & 7.70$^{3}$   & & \\
    &&&&&& 9&39 & 7.96$^{4}$  & & \\
0859$-$039 &23.78& 7.77 & 0.51 &29.1& 21
         & 23&22 & 7.84$^{12}$ & \\
    &&&&&& 24&20 & 7.88$^{18}$  & & \\
1042$-$690 &21.42&7.78 & 0.51 &36.3 &33
         &20&64 & 7.73$^{6}$  & & \\
    &&&&&& 21&38 & 7.86$^{4}$ & & \\
{\bf 1105$-$048} &15.28& 7.83 &  0.52 &24.5 &142
         & 15&58 & 7.81$^{4}$ & $<$32$^{13}$ & \\
    &&&&&& 15&54 & 7.82$^{2}$ & & \\
    &&&&&& 15&87$^{7}$ &      & & \\
{\bf 2359$-$434} &8.66&8.56* & 0.95 & --- &2200
         & 7&76$^{7}$ &       & $\approx$0$^{10}$ & $7.8\pm 0.4$Y$^{15}$ \\
    &&&&&& 8&72 & 8.58$^{4}$ & & \\
    &&&&&& 8&85 & 8.57$^{9}$ & & \\
    &&&&&& 8&67 & 8.83$^{11}$ &&\\
\hline
\end{tabular}
\footnotesize{References: 
$^1$\cite{Bergeron-etal:89};         $^2$\cite{Bergeron-etal:92};   
$^3$\cite{Bergeron-etal:01};         $^4$\cite{Bragaglia-etal:95};
$^5$\cite{Finley-etal:97};           $^6$\cite{Kawka-etal:00}; 
$^7$\cite{Kepler-Nelan:93};          $^8$\cite{Kidder-etal:92};
$^9$\cite{Koester-Allard:93};        $^{10}$\cite{Koester-etal:98};
$^{11}$\cite{Koester-etal:01};       $^{12}$\cite{Napi-etal:99}; 
$^{13}$\cite{Pilachowski-Milkey:84}; $^{14}$\cite{Saffer-etal:88};       
$^{15}$\cite{vanAltena-etal:95};     $^{16}$\cite{Vauclair-etal:97};      
$^{17}$\cite{Vennes:99};             $^{18}$\cite{Vennes-etal:97}.} \\
\footnotesize{$^*$ gravity fixed at a value which reproduced the absolute 
brightness computed from the parallax.} 
\end{center}
\end{table*}

\subsection{Comments on individual objects}

\paragraph{WD\,0135$-$052 (L\,870$-$2):} 
We have only one single observation of this target, which indicates a very 
weak magnetic field  of 650\,G on a $2\sigma$ level (see Sect. 4.2). It is 
reported as a double-lined spectroscopic binary composed of a detached pair 
of DA white dwarfs with an orbital period of the system of 1.556 days 
\citep{Saffer-etal:88}. The system has a mass ratio $q=M_1/M_2=0.90$ 
\citep{Saffer-etal:88,Bergeron-etal:89}, \ie\ the brighter component is less 
massive. The luminosity ratio of both white dwarfs amounts to 1.5.

\paragraph{WD\,0346$-$011 (GD\,50):} 
\cite{Bragaglia-etal:90} list this white dwarf as a suspected double 
degenerate from their radial velocity measurements. However, the more accurate 
radial velocity measurements performed by Maxted et al.\ (2000) and from the 
SPY spectra indicate a constant radial velocity. GD\,50 is a very massive 
white dwarf ($M=1.24M_\odot$). However, our spectropolarimetric measurements 
did not detect a significant magnetic field. In our sample, this is the star 
with the broadest Balmer lines and, consequently, the weakest limit on the 
magnetic field strength.

\paragraph{WD\,0446$-$789:}
We found a magnetic field of 4280\,G in this relatively young white dwarf 
(cooling age of 21 Myr) with a mass ($0.49M_\odot$) slightly lower than the 
peak of the white dwarf mass distribution. \cite{Maxted-etal:00} and our SPY 
observations show that the radial velocity of this star is constant and thus 
membership in a close binary system can be ruled out.

\paragraph{WD\,0612+177 (LTT\,11818):} 
By the time a star reaches $\sim$45000 \kelvin\, the process of H 
accumulation has progressed to the point that all white dwarfs are DA in 
appearance and no DB white dwarfs are observed in the range 30000-45000 
\kelvin, the so-called ``DB Gap''. WD\,0612+177 lies near the red edge of the 
DB Gap \citep{Holberg-etal:90}, where the onset of near surface convection can 
mix He into the photospheres of those stars having the thinnest H layers. 
\cite{Holberg-etal:90} reported the detection of a 
weak feature due to He~{\sc i} $\lambda$4471 in three high signal-to-noise 
spectra of this white dwarf. From the observed strength of these features 
those authors obtained estimates of a homogeneous He/H ratio of 
log (He/H) = -2.56$\pm$0.26 or, alternatively, a stratified H envelope mass 
of log (M$_{\rm H}$/\Msolar) $\sim$ $-$16.6$\pm$0.3. However, subsequent 
optical observations of this WD by \cite{Kidder-etal:92} did not reveal the 
4471 \ang\ features at the strength of the earlier observations. We also do 
not detect this feature in our two single observations of this WD. One 
possible explanation given by Kidder \etal\ is that photospheric He may not 
be distributed uniformly over the surface of WD\,0612+177, possibly due to 
the presence of a magnetic field and a slow stellar rotation which leads to 
a modulation of the He~{\sc i} line strengths. 
Our observations suggest a 3$\sigma$ upper limit of the magnetic field of less 
than 2\,kG from two different observing dates. This rather tight upper limit 
needs to be taken into account when considering scenarios to produce a 
non-uniform distribution of He and H in the white dwarf atmosphere. 

\paragraph{WD\,0839$-$327:} 
The quality of our spectral fit is poor compared to other stars in our 
sample.  Our result is virtually identical to the value derived by 
\citet{Bragaglia-etal:95} using the same technique. \citet{Bergeron-etal:01} 
using optical and infrared broadband photometry combined with a trigonometric 
parallax derived lower temperature and gravity. This might indicate that this 
object is an unresolved binary, consisting of two white dwarfs. This white 
dwarf is reported as a double degenerate binary by \cite{Bragaglia-etal:90}. 
However, the more accurate measurements done for the SPY project indicate a 
constant radial velocity ruling out a very close system.

\paragraph{WD\,1042$-$690:}
This white dwarf has an M type companion in a 0.34 days orbit 
\citep{Kawka-etal:00}. Some contamination by line emission from the cool 
companion (mainly Balmer lines and Ca H+K) is present in our spectrum. The 
contaminated parts of the line profiles were excluded from our fits. 
 Broad-band linear polarization could be measureable for this binary due 
to irradiation and reflection effects. This should, however, not affect our 
measurement of the  circular polarization in spectral lines.

\paragraph{WD\,1105$-$048:} 
For this target we have discovered in the second of two observations the
spectropolarimetric signatures from a magnetic field of $-$3900\,G. 
WD\,1105$-$048 resides in a common proper motion binary with a dM5 companion 
separated by $279''$, corresponding to 6840\,AU \citep{Oswalt-etal:88}. 
The mass of this white dwarf ($0.52M_\odot$) is very close to the peak
of the white dwarf mass distribution. 

\paragraph{WD\,2359$-$434 (L\,870$-$2):}
For WD\,2359$-$434 we have detected a magnetic field of $-$3100 G. Already 
\cite{Koester-etal:98} reported a very shallow and narrow core of the 
H$\alpha$ profile observed in this object. They speculated that what they see 
is only the unshifted component of a Zeeman triplet, with the other components 
shifted outside the observed spectral range or smeared out due to the 
inhomogeneity of the field. In favour of this speculation is the unusually 
high surface gravity of this object, since magnetic white dwarfs tend to be 
more massive than non-magnetic objects \citep{Liebert:88}. However, it would 
require a magnetic field $>50$\,kG to shift the $\sigma$ components outside 
the spectral range of the Koester et al. observation. No further components 
could be detected in our spectra or the SPY spectra.

We noticed during the fitting process that our FORS data show flat Balmer 
lines cores. Available UVES spectra of this white dwarf show that the core of 
H$\alpha$ is also flat \citep[as already reported by][]{Koester-etal:98}. 
Since an accurate parallax measurement exists for WD\,2359$-$434, we modified
our fitting procedure for this object. The gravity was fixed at a value, which 
reproduced the absolute brightness of the white dwarf computed from the 
parallax. The temperature was determined from a fit of the line wings of the 
Balmer lines, excluding the cores. Since the absolute brightness depends on 
the stellar temperature as well, a few iterations were necessary, before we 
got a self-consistent solution. The derived temperature (8.66\,kK) is 
consistent with the result of \citet{Koester-Allard:93}, who fitted UV spectra 
taken with IUE. \citet{Kepler-Nelan:93} using the same technique, derived a
lower temperature (7.76\,kK). This is possibly caused by the low value of the 
surface gravity ($\log g = 8.0$) adopted during their fitting procedure, which 
is much lower than plausible values for WD\,2359$-$434.

The reason for the flat Balmer line cores of WD\,2359$-$434 remains a mystery. 
Being a white dwarf with a broad range of magnetic fields is a possible 
explanation since it would smear out the sigma components. However, in our 
polarization measurements there is no indication for such a broad range.

\section{Discussion and conclusions}
In this work we have used the spectropolarimetric capability of the FORS1 
instrument, together with the light collecting power of the VLT, in order to 
investigate the presence of magnetic fields in the range $1-10$~kG for a 
sample of 12 white dwarfs. Three of the stars out of 12 normal DA white dwarfs 
of our sample, WD\,0446-789, WD\,1105-048 and WD\,2359-434, exhibit magnetic 
fields of a few kilogauss in one or all available observations (see Figs.\,3, 
4 and 5). The detection rate of 25~\% suggests now strongly that a substantial 
fraction of white dwarfs have a weak magnetic field. 
 
With the exception of the bright white dwarf 40 Eri B, for which a magnetic 
field of only 4\,kG had earlier been detected \citep{Fabrika-etal:03}, 
WD\,0446$-$789, WD\,1105$-$048 and WD\,2359$-$434 have the weakest magnetic 
fields detected so far in white dwarfs. In previous extended searches for weak 
fields in white dwarfs \citep[\eg][]{Schmidt-Smith:95} only very few objects 
have been found. Only six detections have been reported for magnetic fields 
below 100\,kG, which are not all confirmed, and only three objects have a 
field weaker than about 50\,kG. \cite{Kawka-etal:03} have reported 
longitudinal magnetic fields in three stars but no significant detection was 
made because their $1\sigma$ error was almost as large as the observed value 
itself. They concluded that the population of white dwarfs with magnetic 
fields in excess of 1\,MG is well known, but that lower-field white dwarfs 
remained undetected. 

Note, however, that our investigation is based on the averaged longitudinal
component of the magnetic field, meaning that the maximum magnetic field at 
the white dwarf surface can be stronger, depending on the field geometry
(described \eg\ by offset dipoles, or more complex distributions; with the 
underestimate being larger for a more complex magnetic distribution) and on 
the orientation relative to the observer. Therefore, our results for the three 
objects with a positive detection are lower limits, since cancellation effects 
are expected.

The population of known magnetic white dwarfs presently comprises some 125 
stars \citep{Wickramasinghe-Ferrario:00,Gaensicke-etal:02,Schmidt-etal:03}.
\cite{Wickramasinghe-Ferrario:00} and \cite{Jordan:01} established that about 
3\% of all white dwarfs had a magnetic field above 100\,kG. Results from 
the first two years of the Sloan Digital Sky Survey 
\citep{Gaensicke-etal:02,Schmidt-etal:03} show the total number of 
known magnetic white dwarfs with $B \geq 1.5$ MG being 6\%. 
Recently, \cite{Liebert-etal:03} have found that the incidence of magnetism at 
the level of $\sim$ 2\,MG or greater is at least $\sim$10\%, or higher. They 
suggest that the total fraction of magnetic WDs may be substantially higher 
than 10\% due to the limited spectropolarimetric analyses capable of detecting 
lower field strengths down to $\sim$ 10\,kG.
Our 3 detections out of 12 objects seem to indicate that low magnetic fields 
on white dwarfs ($<$10\,kG) are frequent while high magnetic fields are 
relatively rare. However, with only three detections this hypothesis remains 
insecure. If confirmed by future observations, the investigation of weak 
magnetic fields in white dwarfs could form a cornerstone for the future 
investigation of the properties and evolution of stellar magnetic fields.

Our sample of white dwarfs is too small to discuss in detail the dependence 
of the magnetic field strength on the stellar parameters (masses and cooling 
ages). It is, however, worthwhile to mention that two of our detections 
(WD\,0446$-$789 and WD\,1105$-$048) have masses of only 0.5 \Msolar. This 
means that their progenitors on the main-sequence had less than 1 \Msolar\ 
\citep{Weidemann:2000}. These two stars are therefore very different from the 
majority of white dwarfs with megagauss magnetic fields which tend to have 
higher masses \citep{Greenstein-Oke:82,Liebert:88} and, therefore, high-mass 
parent stars. 

Measurements of weak magnetic fields are now possible for many white dwarfs 
with the new large telescopes, which allow a magnetic field function (MFF, in 
analogy to the mass function) to be constructed in the $1-100$~kG range once a
sufficient number of detections have been made. 
Such a MFF can be compared to the corresponding function for 
main-sequence stars \citep{Bychkov-etal:97} and will provide input for answers 
to the following key questions on the evolution of magnetic fields in stars: 
Are the magnetic fields in white dwarfs just the fossil relics of magnetic 
main-sequence stars strengthened by contraction due to conservation (to a 
large extent) of magnetic flux? Or do the magnetic  fields develop 
considerably through
the final stages of stellar evolution? Are the strongly magnetic white 
dwarfs a distinctive class of objects or do they just represent a tail of the 
distribution of magnetic fields present in all white dwarfs? Is there a 
dependence between magnetic field strength and mass as found in the case of 
magnetic WDs with higher field strengths? Do the magnetic field strengths 
correlate with temperature, which would be a hint for a decay on the white 
dwarf cooling sequence?

 Several authors have suggested that the frequency of magnetic white 
dwarfs may increase with decreasing effective temperature, luminosity and 
with increasing cooling age 
\citep[\eg\ ][]{Valyavin-Fabrika:98,Liebert-etal:03}, and may decrease 
sharply with distance \citep[][]{Fabrika-Valyavin:98}.

Alternatively to the fossil origin of the magnetic field in white dwarfs,
\cite{Markiel-etal:94} and \cite{Thomas-etal:95} have shown that a weak 
magnetic field of $\approx 1.3$\,kG in the variable DB star GD\,358 can be 
explained by an $\alpha\omega$ dynamo. The magnetic field in this star has 
been inferred indirectly by analyzing the $g$-mode oscillation spectrum taken 
with the WET \citep[Whole Earth Telescope,][]{Winget-etal:94}. However, 
according to the atmospheric parameters, the convection zone in all of our 
sample stars should be too shallow to support an $\alpha\omega$ dynamo.

Another long-standing problem in white dwarf research is the issue why metals 
are accreted by helium-rich white dwarfs in the range $8000\,\kelvin < 
T_{\rm eff} < 15000$\,\kelvin\ during the passage through an interstellar 
cloud while almost no hydrogen is brought into the white dwarf atmosphere. 
\cite{Illarionov-Sunyaev:75} suggested that fields below $10^5$\,G provide a 
screening mechanism to separate hydrogen and ionized species from grains in 
white dwarfs accreting from the interstellar matter. Since this phenomenon 
always occurs in this type of star, it is possible that at this level all 
white dwarfs contain magnetic fields. \cite{Friedrich:04} have searched for 
circular polarisation in one DBZ and one DBAZ which have accreted metals, but 
three or four orders of magnitude less hydrogen than expected. 
In one case (L745-46A) a magnetic field of 7\,kG (1 $\sigma$ error of
$\pm 2$\,kG; 99\%\ confidence interval of $\pm 6$\,kG) was found, which, 
however, was based on the H$\alpha$ line only. For the second object (GD\,40), 
only an upper limit of 12\,kG (99\% confidence) could be derived from
polarization measurements around three spectral lines. Theoretically,
magnetic fields of 4\,kG and 250\,kG, respectively, would be required for 
the screening by the propeller mechanism to be efficient around the
two stars.

At our signal-to-noise ratio, magnetic fields down to about 2 \,kG can be 
measured. There is still the possibility that all magnetic white dwarfs 
contain surface magnetic fields at the 1\,kG level. In order to test this 
hypothesis, much longer exposure times would be necessary, even with the VLT. 

\acknowledgements{
We gratefully acknowledge useful comments on the effective Land{\'e} factor by 
E. Landi degl'Innocenti. We acknowledge the use of LTE model spectra computed 
by D. Koester. We thank the staff of the ESO VLT for carrying out the service 
observations. Work on magnetic white dwarfs in T{\"u}bingen is supported by the 
DLR grant 50 OR 0201. R.N. acknowledges support by a PPARC Advanced Fellowship.
We would like to thank the referee J. D. Landstreet for valuable suggestions.}

\bibliographystyle{aa}

\end{document}